# A bottle of tea as a universal Helmholtz resonator

Martín Monteiro[a], Cecilia Stari[b], Cecilia Cabeza[c], Arturo C. Marti[d],

[a] Universidad ORT Uruguay; monteiro@ort.edu.uy

[b] Universidad de la República, Uruguay, cstari@fing.edu.uy

[c] Universidad de la República, Uruguay, cecilia@fisica.edu.uy

[d] Universidad de la República, Uruguay, marti@fisica.edu.uy

Resonance is an ubiquitous phenomenon present in many systems. In particular, air resonance in cavities was studied by Hermann von Helmholtz in the 1850s. Originally used as acoustic filters, Helmholtz resonators are rigid-wall cavities which reverberate at given fixed frequencies. An adjustable type of resonator is the so-called *universal Helmholtz resonator*, a device consisting of two sliding cylinders capable of producing sounds over a continuous range of frequencies. Here we propose a simple experiment using a smartphone and normal bottle of tea, with a nearly uniform cylindrical section, which, filled with water at different levels, mimics a universal Helmholtz resonator. Blowing over the bottle, different sounds are produced. Taking advantage of the great processing capacity of smartphones, sound spectra together with frequencies of resonance are obtained in real time.

**Helmholtz resonator**

Helmholtz resonators consist of rigid-wall containers, usually made of glass or metal, with volume $V$ and neck with section $S$ and length $L$ [1] as indicated in the Fig. 1. In the past, they were used as acoustic filters, for the reason that when someone blows over the opening, air inside the cavity resonates at a frequency given by

$$f = \frac{c}{2\pi}\sqrt{\frac{A}{VL'}} .$$

In this expression $c$ is the sound speed and $L'$ is the equivalent length of the neck, accounting for the end correction, which in the case of outer end unflanged, results $L' = L + 1.5a$. The previous expressions are valid provided that the linear dimensions are smaller than the typical wavelength ($L \ll \lambda, a \ll \lambda, h \ll \lambda$). A universal Helmholtz resonator is a special type of resonator in which the volume can be varied resulting in a range of possible frequencies of resonance.

**A smartphone-based experiment in acoustic**

There are several examples of physics experiment using smartphones in the field of acoustic [2-8]. In some of them the smartphone is used as a microphone to digitalize sound or as a source of pure or more complex sounds[1-3]. An important advantage of smartphones, not always taken into account, is their capacity to process speedily a lot of information, and in particular, to obtain sound spectra in real time [4-8].

To obtain the volume of the cavity, the experiments starts placing the empty bottle on a weighing scale and setting it to zero. After that, the bottle is filled with water until the neck (7.0 cm below the opening) and the reading in the scale, in this case, 435 g, indicates the cavity volume, that is, $V = 435$ cm$^3$.

Next, the bottle is filled with water at different levels. For each water level, the volume of the air cavity is obtained subtracting the added water to the total volume of the cavity. Blowing across the top of the bottle, with the lower lip touching the edge of the bottle produces a resonance in the cavity. The Advanced spectrum *app* is used to obtain the spectrum in real time and the highest peak which corresponds to the resonance frequency as shown in Fig. 2. The "hold" button is useful to frozen the display and write down the frequency. So, varying the volume of water in the bottle it is possible to found the relationship between volume and resonance frequency.

**Results and analysis**

In Fig. 3 we show the linearized relationship between the resonance frequency and the cavity volume, specifically, we plot the $f^2(1/V)$. Note that corresponding wavelengths, between 1 and 1.5 m are much larger than the linear dimensions of the cavity.

The sound speed c is related to the slope as $\dfrac{c^2 A}{4\pi^2(L+1.5A)}$. In our experiment, the sound speed obtained is $c = 345(5)$ m/s resulting in an uncertainty of less than 2%. Taking into account that the ambient temperature was 22.3 ºC, the value exhibits a great coherence with reference value, 344 m/s, at this temperature. We note that there are at least a couple of uncontrollable sources of uncertainties in this experience: the temperature of the blown air is a bit larger than the ambient temperature and the contour conditions at the opening are modified by the blow and the proximity of the lips.

To sum up, we presented a very simple experiment using everyday stuff to study acoustic resonance. This experiment is based on the great processing capacity of smartphones to obtain real time spectra. The result obtained for the sound speed is in concordance with the standard value.

**References**

[1] Kinsler, L. E., Frey, A. R., Coppens, A. B., & Sanders, J. V. (1999). *Fundamentals of Acoustics, 4th Edition, Wiley-VCH.*
[1] Kuhn, J., & Vogt, P. (2013). "Analyzing acoustic phenomena with a smartphone microphone" *The Physics Teacher*, 51(2), 118-119.

[2] Yavuz, A. (2015) "Measuring the speed of sound in air using smartphone applications" *Physics Education*, 50(3), 281.

[3] Kasper, L., Vogt, P., & Strohmeyer, C. (2015). "Stationary waves in tubes and the speed of sound" *The Physics Teacher*, 53(1), 52-53.

[4] Parolin, S. O., & Pezzi, G. (2013). "Smartphone-aided measurements of the speed of sound in different gaseous mixtures" *The Physics Teacher*, 51(8), 508-509.

[5] Monteiro, M., Martí, A. C., Vogt, P., Kasper, L., & Quarthal, D. (2015). "Measuring the acoustic response of Helmholtz resonators" The Physics Teacher, 53(4), 247-249.

[6] Hirth, M., Kuhn, J., & Müller, A. (2015). "Measurement of sound velocity made easy using harmonic resonant frequencies with everyday mobile technology" *The Physics Teacher*, 53(2), 120-121.

[7] González, M. and González, M. (2016). "Smartphones as experimental tools to measure acoustical and mechanical properties of vibrating rods" *European Journal of Physics*, 37(4), 045701

[8] R. Jaafar, S. Kadri Ayop, A. Tarmimi, K. Keng Hon, A. Nazihah Mat Daud, and M. Helmy Hashim (2016). "Visualization of Harmonic Series in Resonance Tubes Using a Smartphone" *The Physics Teacher* 54(9), 545 – 547.

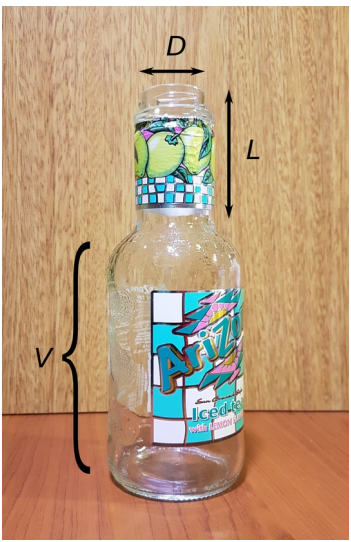

**Figure 1.** The tea bottle, with a nearly uniform circular section, is suitable for the present experience and dimensions: d=29,0(2) mm and L=70(1) mm.

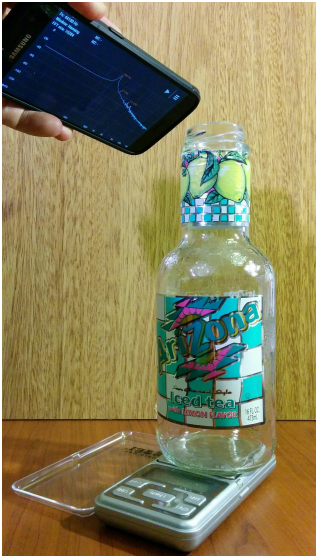

**Figure 2.** Experimental setup: the bottle on the scale and the smartphone displaying a sound spectrum.

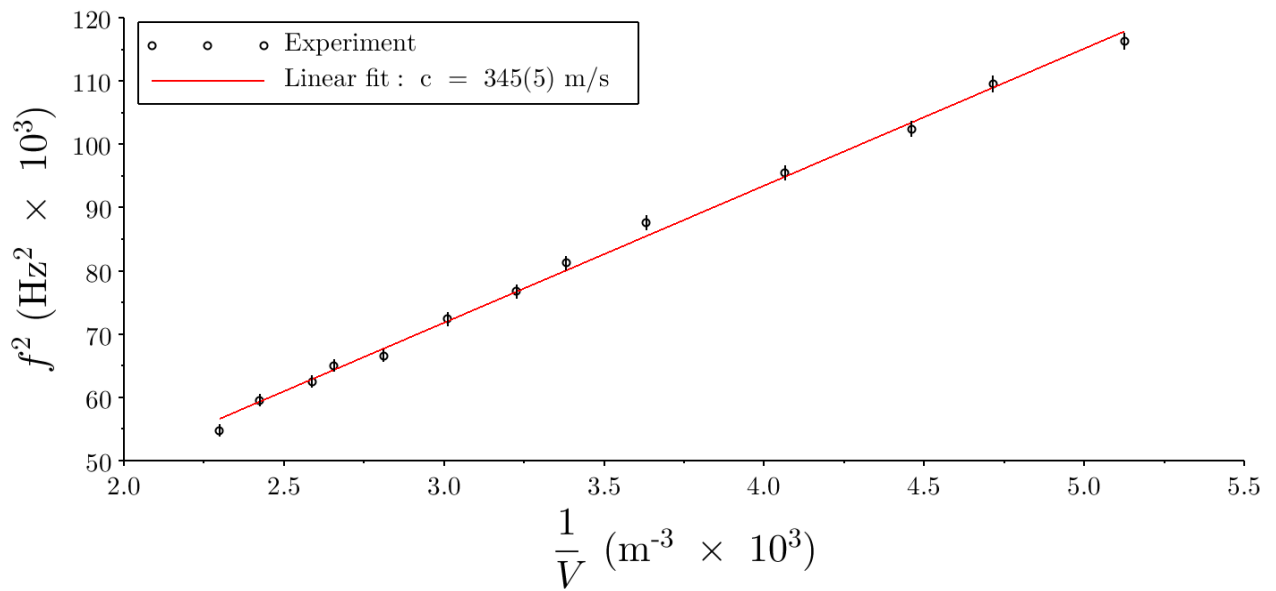

**Figure 3.** Linearized relationship between resonance frequency and air volume: experimental values and linear fit.